\documentclass[prd,amsmath,amssymb,twocolumn,nofootinbib]{revtex4}

\usepackage{epsfig}
\usepackage{multirow}

\newcommand{\PRE}[1]{}
\newcommand{\be}{\begin{equation}}
\newcommand{\ee}{\end{equation}}
\newcommand{\bea}{\begin{eqnarray}}
\newcommand{\eea}{\end{eqnarray}}
\newcommand{\nn}{\nonumber}
\newcommand{\ba}{\begin{array}}
\newcommand{\ea}{\end{array}}

\newcommand{\lsim}{\mathrel{\hbox{\rlap{\hbox{\lower4pt\hbox{$\sim$}}}\hbox{$<$}}}}
\newcommand{\gsim}{\mathrel{\hbox{\rlap{\hbox{\lower4pt\hbox{$\sim$}}}\hbox{$>$}}}}
\newcommand{\til}{\widetilde}

\newcommand{\tev}{\text{TeV}}
\newcommand{\gev}{\text{GeV}}

\newcommand{\md}{\text{~mod~}}
\newcommand{\br}{{\tt Br}}
\newcommand{\fb}{\text{fb}}
\newcommand{\ps}{$~~$}
\newcommand{\ms}{-}

\begin{document}

\title{
\PRE{\vspace*{1.5in}}
Minimal gauge origin of baryon triality and flavorful signatures at the LHC
}
\author{Hye-Sung Lee}
\affiliation{Department of Physics, Brookhaven National Laboratory, 
Upton, NY 11973, USA}

\begin{abstract}
\PRE{\vspace*{.1in}} \noindent
Baryon triality ($B_3$) is a $Z_3$ discrete symmetry that can protect the proton from decay.
Although its realization does not require supersymmetry, it is particularly appealing in the supersymmetry as an alternative to the popular $R$-parity.
We discuss the issues in gauging $B_3$, and present the minimal supersymmetric model with $B_3$ as the remnant discrete symmetry of a TeV scale $U(1)$ gauge symmetry.
A flavor-dependent $U(1)$ charge is necessary to achieve this, and it results in very distinguishable and flavorful predictions for the LHC experiments.
We find a complementarity between a 2-lepton sneutrino resonance and a 4-lepton $Z'$ resonance in the supersymmetry search when a certain condition is satisfied.

In addition, we introduce baryon tetrality ($B_4$), which would play an equivalent role if there are four fermion generations.
\end{abstract}

\maketitle

\section{Introduction}
\label{sec:intro}
While the standard model (SM) of particle physics has been extremely successful in explaining data, it also casts a puzzle associated with the quantum correction of the Higgs boson mass.
Supersymmetry (SUSY) is a prevailing new physics paradigm that can address the issue elegantly, yet a mere supersymmetrization of the SM creates other problems.
Some problems such as fast proton decay originate from the fact that the lepton number ($L$) or baryon number ($B$) violations occur at the renormalizable level unlike the SM.
This is because of the scalar fields which are superpartners of the SM fermions in the SUSY.

An additional mechanism to control $L$/$B$ violations is a categorical requirement of any SUSY models.
While a discrete symmetry is presumably the simplest option, it is preferred to be accompanied by a gauge origin to avoid any potential breakdown by Planck scale physics \cite{Krauss:1988zc}.

A typical example of the discrete symmetry from a $U(1)$ gauge symmetry in SUSY is $R$-parity (equivalently, matter parity) from $U(1)_{B-L}$ gauge symmetry \cite{Font:1989ai,Martin:1992mq}:
\be
U(1)_{B-L} ~ \to ~ R_2 .
\ee
$R$-parity forbids all renormalizable level $L$/$B$ violating operators and protects the proton from them.
Another merit of the $R$-parity is that it makes the lightest supersymmetric particle (LSP) stable, providing a dark matter candidate.

The minimal supersymmetric standard model (MSSM) which uses $R$-parity has been extensively exploited in the literature as a TeV scale SUSY model, and the most experimental search schemes for the SUSY are based on this model. (For a review, see Ref.~\cite{Martin:1997ns}.)
While the MSSM remains as a charming possibility, it is compulsory to investigate other possibilities for the TeV scale SUSY model.
After all, the MSSM is not the only possibility and the best search strategy of the SUSY as well as the associated Higgs and dark matter search strategies at the Large Hadron Collider (LHC) experiments in other models can be different.

In this paper, we present a SUSY model with $B_3$ (baryon triality) as the {\it only} remnant discrete symmetry from a $U(1)$ gauge symmetry with the minimal particle spectrum:
\be
U(1)_{B-x_i L} ~ \to ~ B_3 .
\ee
As we will show later, when the minimal particle spectrum is assumed, family-dependent $U(1)$ charges are required.
(See Ref.~\cite{Langacker:2000ju} for a general discussion about the family-dependent $U(1)$ charges.)

When an Abelian gauge symmetry is introduced in the SUSY framework, its typical scale is TeV scale if it has ordinary size of couplings.
The masses of scalar fermions get an extra $D$-term contribution from the $U(1)$ gauge symmetry.
If the SUSY is the solution to the gauge hierarchy problem, the scalar fermions (such as scalar top) should not be much heavier than the electroweak or TeV scale.
Thus the scale of a new $U(1)$ gauge symmetry or its gauge boson ($Z'$) should not exceed TeV scale unless there is a cancellation mechanism.
The exceptions to this argument can occur when the $U(1)$ symmetry breaking is associated with an $F$ and $D$ flat directions.
(For the detailed discussion on the $U(1)$ symmetry breaking scales, see Sec.~III in Ref.~\cite{Langacker:2008yv} and references therein.)

Since a light $Z'$ has not been discovered, we expect it to be likely around TeV scale which is within the scope of the LHC experiments.
A TeV scale $Z'$ combined with the unique features of the new symmetry can lead to very distinguishable predictions for the LHC collider experiments, as we discuss briefly later in this paper.

In addition, we discuss the case when the fermion generation is four to reflect recent ample interest in the 4th-generation scenarios.
We introduce $B_4$ (baryon tetrality) from
\be
U(1)_{B-x_i L} ~ \to ~ B_4 ,
\ee
which can play the similar role of the $B_3$ in the 4th-generation scenario.

The outline of the rest of this paper is given as following.
In Section~\ref{sec:B3}, we describe the $B_3$.
In Section~\ref{sec:conditions}, we describe the conditions for a $U(1)$ gauge symmetry to have only the $B_3$ as a residual discrete symmetry.
In Section~\ref{sec:others}, we briefly discuss the dark matter and other issues in our model where the $R$-parity is absent.
In Section~\ref{sec:LHC}, we qualitatively describe how the LHC phenomenology of our model can be different from the typical SUSY search.
In Section~\ref{sec:B4}, we describe the $B_4$ for the 4th-generation scenario.

\section{Baryon triality}
\label{sec:B3}
$B_3$ is a $Z_3$ discrete symmetry suggested by Ibanez and Ross in 1992 as an alternative to $R$-parity \cite{Ibanez:1991pr}.
It can avoid dimension 5 proton decay operators (such as $QQQL$ and $U^cU^cD^cE^c$), which are allowed by $R$-parity.
Numerous theoretical and phenomenological studies for the models based on $B_3$ have followed.
(For examples, see Refs.~\cite{Castano:1994ec,Babu:2003qh,Dreiner:2005rd,Dreiner:2006xw,Lee:2007qx,Lee:2008pc,Bernhardt:2008jz,Lee:2008zzl}.)

The discrete charge of $B_3$ is
\be
B_3 :~ q = - B + 2 Y \md 3
\ee
where $Y$ is hypercharge.
(See Table~\ref{tab:charges} for the charges.) 
This establishes a selection rule $\Delta B = 3 \times \text{integer}$, under which $B$ can be violated only by $3 \times \text{integer}$ while $L$ can be freely violated.
Thus the proton decay ($\Delta B = 1$ process) never occurs while the $L$ violating processes can occur at the renormalizable level.

It is well known that $B-L$ is the only anomaly-free $U(1)$ gauge symmetry unless exotic fields charged under the SM gauge groups are added.
A caveat is that this statement is true only for the family-independent $U(1)$ charges.
Other $U(1)$ symmetries are possible when we allow family-dependent $U(1)$ charges.
Besides, $U(1)_{B-L}$ cannot have the $B_3$ as its remnant discrete symmetry.

Thus, there are two directions in gauging $B_3$: (i) to allow exotic fermions charged under the SM gauge groups, (ii) to allow family-dependent $U(1)$ charges.

Refs.~\cite{Lee:2007qx,Lee:2008zzl} dealt with the exotic fermions case, keeping the MSSM sector charges family-independent.
In this paper, we keep the minimal particle contents, and generalize $U(1)$ charges from $B-L$ to $B-x_i L$ ~($x_i$: family-dependent value) without adding exotic fields.
By the {\it minimal} particle spectrum, we mean the same particle spectrum as the usual supersymmetric $B-L$ model: MSSM fields (including 3 right-handed neutrinos) plus an additional gauge boson ($Z'$) and two Higgs singlets ($S_1$, $S_2$).
Two Higgs singlets with opposite charges can cancel the anomaly as two Higgs doublets in the MSSM sector do.

Under $B-x_i L$, the $U(1)$ charges are
\bea
&& z[Q] = -z[U^c] = -z[D^c] = B = 1/3 \\
&& z[L_i] = -z[N_i^c] = -z[E_i^c] = -x_i L = -x_i \\
&& z[H_u] = z[H_d] = 0
\eea
and $z[S_1] = -z[S_2]$ should be determined by the condition for the $U(1)$ to have a $Z_3$ as its total remnant discrete symmetry as we discuss in the following section.

The $B-x_i L$ is the same symmetry we used in Ref.~\cite{Lee:2010hf}.
There, the goal was to take $B-x_i L$ as a gauge origin of the $R$-parity.
Here, we take it as a gauge origin of $B_3$, which we regard as the {\it only} remnant discrete symmetry. (Thus $R$-parity is not conserved\footnote{For a review of $R$-parity violation, see Ref.~\cite{Barbier:2004ez}. For the recent works in the spontaneous $R$-parity violation, see Refs.~\cite{FileviezPerez:2008sx,Barger:2008wn,FileviezPerez:2009gr,Everett:2009vy}.}.)

\begin{table}[tb]
\begin{tabular}{|l||ccc|cc|cc|}
\hline
          & ~$B\!-\!x_i L$ & ~$6Y$ & ~$(B\!-\!x_i L)\!-\!2Y$ & ~$B_3$ & ~$B_4$ & ~$B_N$ & $L_N$ \\
\hline
~$Q$~     & $\ps 1/3$    & $\ps 1$ & $\ps 0$      & $\ps 0$ & $\ps 0$ & $\ps 0$ & $\ps 0$ \\
~$U^c$~   & $\ms 1/3$    & $\ms 4$ & $\ps 1$      & $\ms 1$ & $\ms 1$ & $\ms 1$ & $\ps 0$ \\
~$D^c$~   & $\ms 1/3$    & $\ps 2$ & $\ms 1$      & $\ps 1$ & $\ps 1$ & $\ps 1$ & $\ps 0$ \\
~$L_i$~   & $\ms x_i$    & $\ms 3$ & $\ps 1-x_i$  & $\ms 1$ & $\ms 1$ & $\ms 1$ & $\ms 1$ \\
~$N_i^c$~ & $\ps x_i$    & $\ps 0$ & $\ps x_i$    & $\ps 0$ & $\ps 0$ & $\ps 0$ & $\ps 1$ \\
~$E_i^c$~ & $\ps x_i$    & $\ps 6$ & $\ps x_i-2$  & $\ms 1$ & $\ps 2$ & $\ps 2$ & $\ps 1$ \\
~$H_u$~   & $\ps 0$      & $\ps 3$ & $\ms 1$      & $\ps 1$ & $\ps 1$ & $\ps 1$ & $\ps 0$ \\
~$H_d$~   & $\ps 0$      & $\ms 3$ & $\ps 1$      & $\ms 1$ & $\ms 1$ & $\ms 1$ & $\ps 0$ \\
\hline
~$S_1$~   & $\ps z[S_1]$ & $\ps 0$ & $\ps z[S_1]$ &         &         &         &         \\
~$S_2$~   & $\ms z[S_1]$ & $\ps 0$ & $\ms z[S_1]$ &         &         &         &         \\
\hline
\end{tabular}
\caption{Various $U(1)$ charges and discrete charges.}
\label{tab:charges}
\end{table}

\section{Conditions for the minimal model of $B_3$}
\label{sec:conditions}
We need to discuss the conditions for our $B-x_i L$ model with the minimal particle contents (i) to be anomaly-free, and (ii) to have $B_3$ as the remnant discrete symmetry.

For example, $[SU(2)_L]^2$-$U(1)$ anomaly-free condition (with $N_f$ being the number of family) is
\bea
&&N_C \sum_{i=1}^{N_f} z[Q] + \sum_{i=1}^{N_f} z[L_i] + (z[H_u] + z[H_d]) \nn \\
&&= 3 \times (N_f \times 1/3) - \sum_{i=1}^{N_f} x_i + (0+0) \nn \\
&&= 0 .
\eea
The model is completely anomaly-free (for $N_f = 3$) if
\be
x_1 + x_2 + x_3 = 3 , \label{eq:xSum}
\ee
which is required by $[SU(2)_L]^2$-$U(1)$ and $[U(1)_Y]^2$-$U(1)$ anomalies, while the other anomalies are automatically satisfied \cite{Lee:2010hf}.
It is clear that $B-L$ is the only family-independent choice (i.e. $x_1 = x_2 = x_3$) in the $B-x_i L$.

Now we consider the conditions for our $U(1)_{B-x_i L}$ to have the $B_3$ as its only remnant discrete symmetry.

When a $U(1)$ gauge symmetry is spontaneously broken by a vacuum expectation value (vev) of a Higgs singlet $S$, it leaves $Z_N$ as a remnant discrete symmetry in general.
\be
U(1) ~ \to ~ Z_N
\ee

The relation between the $Z_N$ and its $U(1)$ gauge origin is given as followings, after taking the integral $U(1)$ charges (which are not common multiples) by appropriate hypercharge shift and normalization.
\begin{enumerate}
\item $N = | z[S] |$ ~ (or G.C.D. for multiple Higgs singlets)
\item $q[F_i] = z[F_i] \md N$
\end{enumerate}
The $q[F_i]$ and $z[F_i]$ are the $Z_N$ and $U(1)$ charges, respectively, for each field $F_i$.

These conditions basically require every permitted operator whose total $U(1)$ charge is zero should also keep the total discrete charge zero (up to $\md N$), even after all the Higgs singlets that break the $U(1)$ gauge symmetry are replaced by their vevs.

The $U(1)$ charges normalized to be integral since $Z_N$ charges are integers.
To keep the $Z_N$ irreducible, the integral $U(1)$ charges should not be common multiples, and since the hypercharge is independently conserved, it should be made sure that they are not common multiples after any hypercharge shift.

A complete set of possible $Z_N$ discrete symmetry generators for the MSSM fields was first identified in Ref.~\cite{Ibanez:1991pr}: $A_N$, $R_N$, $L_N$.
Among these, $A_N$ is not compatible with the so-called $\mu$-term ($H_u H_d$), which should exist to explain the absence of the very light chargino particle.
In view of this, a generator of the most general $Z_N$ for the MSSM can be written as, using the basis $B_N \equiv R_N L_N$ instead of usual $R_N$ (see Refs.~\cite{Lee:2008pc,Hur:2008sy} for this convention),
\be
g_N = B_N^b L_N^\ell
\ee
with cyclic symmetries $B_N = e^{2\pi i \frac{q_B}{N}}$ and $L_N = e^{2\pi i \frac{q_L}{N}}$ \cite{Ibanez:1991pr}.
The discrete charges of $B_N$ and $L_N$ ($q_B$ and $q_L$, respectively) are shown in Table~\ref{tab:charges}.
The total discrete charge of $Z_N$ is $q = b q_B + \ell q_L \md N$, which can be written as $q = -b B - \ell L + 2 b Y \md N$.

For our $B-x_i L$ model, in order to have a $Z_3$ as a total remnant discrete symmetry, $U(1)$ charges of the Higgs singlets should satisfy (up to overall sign flip)
\be
z[S_1] = -z[S_2] = 3 . \label{eq:zS}
\ee
as the aforementioned Relation 1 requires.

Complying the Relation 2, by equating $(B-x_i L)-2Y \md 3$ to $B_3$ charges up to overall sign (see Table~\ref{tab:charges}), requires $x_i \md 3 = 0$, i.e.
\be
x_i = 3 \times \text{integer} \label{eq:integer}
\ee
in order to have $B_3$ as the remnant discrete symmetry of the $U(1)_{B-x_i L}$.

There are many possible choices of $x_i$ that can satisfy the required conditions:
\be
x_i = (0,0,3), ~(-3,6,0), ~(9,-3,-3), ~\cdots \label{eq:xi} .
\ee

From Eqs.~\eqref{eq:zS} and \eqref{eq:integer}, we can see that some $N^c_i$ can possess the $U(1)$ charges that would allow terms such as $SN^c$ or $SSN^c$, which can induce vev in the scalar component of $N^c$ (right-handed sneutrino) in the absence of $R$-parity.
When $z[N^c_i] = \pm 3$, which is the same as the $z[S]$ required by Eq.~\eqref{eq:zS}, even a more economical particle spectrum without the need of two $S$ fields might be possible in a similar fashion investigated in Ref.~\cite{Barger:2008wn}.

\section{Dark matter and neutrino mixing}
\label{sec:others}
In the absence of $R$-parity, the LSP is not a good dark matter candidate in general.
It is still possible to have a good dark matter candidate without $R$-parity though.
One possible way is to introduce additional hidden sector fields that have charges only under the $U(1)$ gauge symmetry.
It was illustrated for the case of the family-independent $U(1)$ charge case that the hidden sector field can be a stable dark matter candidate, which satisfies the constraints from the relic density measurement and direct detection experiments \cite{Lee:2008pc,Lee:2008em}.

Another dark matter candidate is the gravitino LSP (for instance, see Ref.~\cite{Takayama:2000uz}).
Especially, in the typical gauge mediated SUSY breaking scenario \cite{Giudice:1998bp}, the gravitino is much lighter than the other superpartners due to the hierarchy between the messenger scale and Planck scale.
Because of the small coupling and mass, the gravitino LSP may be still long-lived in the absence of the $R$-parity.

As discussed in Ref.~\cite{Lee:2010hf}, if all $x_i$ are different for all families of leptons, the charged lepton mass matrix and the neutrino mass matrix are both constrained to be diagonal unless additional Higgses charged under the new $U(1)$ are introduced.
This constrained the allowable $x_i$ severely.

Neutrino sector in our model is more complicated due to the contributions from the $L$ violating terms ($\lambda$, $\lambda'$, $\mu'$) \cite{Hall:1983id,Grossman:1998py}.
With the additional $U(1)$ gauge symmetry, there are also possible mixing terms with high dimensional operators \cite{Cleaver:1997nj,Chen:2006hn}.
The high dimensional operators might not provide dominant effect in the collider phenomenology, but because of tiny neutrino masses, they may be relevant in neutrino sector.

As an illustration, let us consider $\nu_e$-$\nu_\tau$ mixing in the $x_i = (-3, 6, 0)$ model (where $z[\nu_e] = 3$ and $z[\nu_\tau] = 0$) through the effective trilinear $\lambda LLE^c$ terms.
They are given by
\bea
&& \lambda_{ijk} L_i L_j E^c_k + \nn \\
&& ~~\sum_{n=1} \left[ \lambda^{(+n)}_{ijk} \left(\frac{S_1}{M}\right)^n + \lambda^{(-n)}_{ijk} \left(\frac{S_2}{M}\right)^n \right] L_i L_j E^c_k \nn \\
&=& \left( \lambda_{ijk} + \lambda^{(+1)}_{ijk} \frac{S_1}{M} + \lambda^{(-1)}_{ijk} \frac{S_2}{M} + \cdots \right) L_i L_j E^c_k
\eea
with some mass parameter $M$, which gives an effective coefficient
\bea
\lambda_{ijk}^\text{eff} = \lambda_{ijk} + \lambda^{(+1)}_{ijk} \frac{\left<S_1\right>}{M} + \lambda^{(-1)}_{ijk} \frac{\left<S_2\right>}{M} + \cdots .
\eea
The $\lambda_{ijk}$, $\lambda^{(+1)}_{ijk}$, $\lambda^{(-1)}_{ijk}$, $\cdots$ cannot coexist.
(If one of them is non-zero, the remaining should be zero because of the invariance under the $U(1)$ gauge symmetry.)
We need two $\lambda^\text{eff}$ to form a loop for the neutrino mixing.
This can be achieved by one $\lambda^\text{eff}$ from $\lambda^{(+1)}_{123} \frac{S_1}{M} L_1 L_2 E^c_3$ term [with total $U(1)$ charge $(3) + (3) + (-6) + (0)$] and the other from $\lambda^{(+2)}_{323} \left(\frac{S_1}{M}\right)^2 L_3 L_2 E^c_3$ term [with total $U(1)$ charge $2 \times (3) + (0) + (-6) + (0)$].
This contribution to the $\nu_e$-$\nu_\tau$ mixing would be absent in the $R$-parity conserving model.

Although some of the solutions in Eq.~\eqref{eq:integer} may not be consistent with neutrino data in the end, detailed analysis to see whether realistic mixings consistent with the experimental data is possible or not would be beyond the scope of this paper.
Besides, details of the neutrino sector may alter if we add more fields or interactions, while keeping the same residual discrete symmetry, to address issues such as the dark matter and strong $CP$ though it might undermine the minimality of the particle spectrum.
(For instance, see Ref.~\cite{Ma:1997nq}.)
Instead of constraining our solutions with specific terms and fields, we find it more useful to discuss the full possibilities of the model at the LHC in the next section putting aside detailed account of the neutrino data\footnote{For a recent review on possible mechanisms for neutrino masses, see Ref.~\cite{Ma:2009dk}.}.

\section{Flavor-sensitive implications for the LHC}
\label{sec:LHC}
Because of the family-dependent $U(1)$ charges in the lepton sector, there are several sources of flavor-dependent leptonic signals predicted for the LHC experiments.
In addition to the flavor-dependent $Z'$ couplings, there are $L$/$B$ violating terms controlled by the $U(1)$.
\bea
S_a H_u L_i &:& \text{(allowed only for $x_i = \pm 3$)} \\
\lambda_{ijk}~ L_i L_j E^c_k &:& \text{(allowed only for $x_k = x_i + x_j$)} \label{eq:lambda} \\
\lambda'_{ijk}~ L_i Q_j D^c_k &:& \text{(allowed only for $x_i = 0$)} \\
\lambda''_{ijk}~ U^c_i D^c_j D^c_k &:& \text{(all forbidden)}
\eea
Further, SM gauge invariances enforce certain antisymmetries in the $L/B$ violating couplings: $\lambda_{ijk} = -\lambda_{jik}$, $\lambda''_{ijk} = -\lambda''_{ikj}$.

Most of the experimental bounds on the $L$/$B$ violating coefficients are given in terms of the ratio to the superpartner masses.
Though constrained by various experiments, quite sizable couplings are allowed.
For superpartner masses at a few$\times 100 ~\gev$, universal $L$ violating couplings (that is, the non-trivial smallest in all combinations of $ijk$) as large as ${\cal O}(|\lambda|) \sim 10^{-3}$ and ${\cal O}(|\lambda'|) \sim 10^{-4}$ are allowed.
The constraints for individual $\lambda_{ijk}$ and $\lambda'_{ijk}$ (that is, for specific choices of $ijk$) can be much weaker.
For extensive table of bounds and the relevant experiments, see Ref.~\cite{Barbier:2004ez}.

In the rest of this section, we discuss some of the flavor-sensitive predictions of our model for the LHC experiments.
We limit our discussion only to the qualitative aspect such as which channels are forbidden while which channels are allowed.
If we specify the gauge coupling constant and masses of particles, we can get quantitative constraints on the model from the $e$-$\mu$-$\tau$ universality \cite{Amsler:2008zzb}, LEP contact interaction constraint \cite{:2003ih}, and Tevatron dilepton resonance searches \cite{Aaltonen:2008vx,Aaltonen:2008ah} and more.
One can find some representative results of the $U(1)_{B-x_i L}$ in Ref.~\cite{Lee:2010hf} for a specific set of parameter values.
When each $x_i$ is different from one another, the SUSY flavor changing neutral current in the lepton sector (such as $\mu \to e \gamma$, $\mu \to e \bar e e$) can be forbidden by the $U(1)_{B-x_i L}$ gauge symmetry \cite{Lee:2010hf}.

\begin{figure}[t]
\begin{center}
\includegraphics[width=0.235\textwidth]{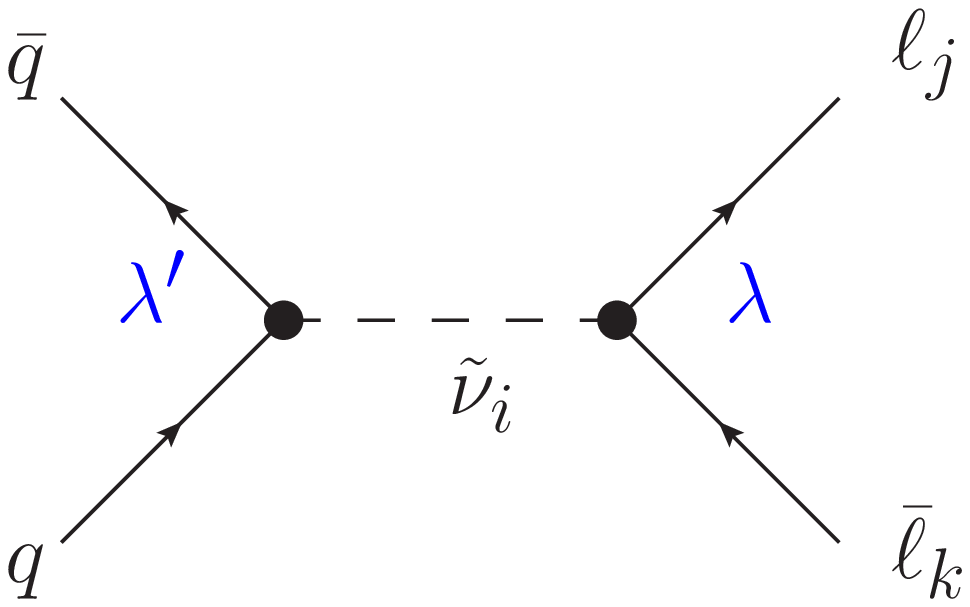}
\includegraphics[width=0.235\textwidth]{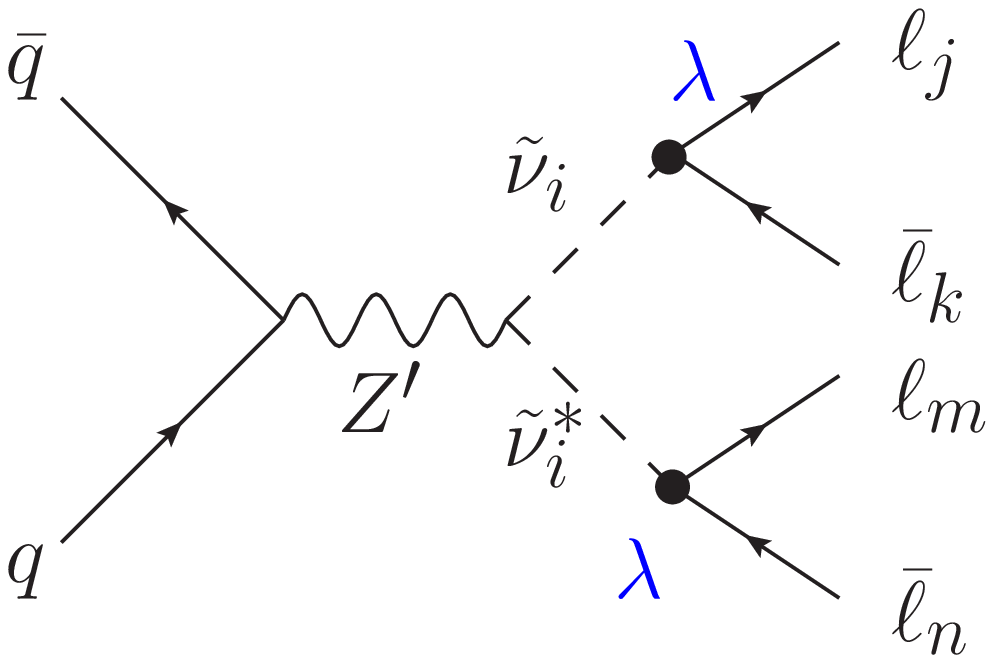} \\
(a) ~~~~~~~~~~~~~~~~~~~~~~~~~~~~~~~~~ (b)
\end{center}
\caption{Complementarity between (a) the 2-lepton $\widetilde \nu$ resonance (with $z[\widetilde \nu] = 0$) and (b) the 4-lepton $Z'$ resonance (with $z[\widetilde \nu] \ne 0$) in the SUSY search.}
\label{fig:diagrams}
\end{figure}

\vspace{0.5cm}
{\it (i) Dilepton $Z'$ resonance at the LHC:}
Dilepton $Z'$ resonance is one of the most direct consequences of any $U(1)$ gauge symmetry.
As discussed in Section~\ref{sec:intro}, the natural scale of $Z'$ is TeV scale in the SUSY framework.

The $Z'$ coupling to lepton is flavor-sensitive in the $U(1)_{B-x_i L}$.
\begin{itemize}
\item $\br(Z' \to e^+ e^-) : \br(Z' \to \mu^+ \mu^-) = x_1^2 : x_2^2$
\end{itemize}
This means the first $Z'$ discovery may depend on the lepton flavor.

For instance, with $x_i = (-3, 6, 0)$, the $e e$ and $\mu \mu$ event rate ratio via the $Z'$ resonance is $1/4$.
The $Z'$ would be discovered by the muon detector much earlier than by the electromagnetic calorimeter at the LHC experiments.
There is no $\tau \tau$ resonance in this example.

The SM background is negligible when we select only the events whose invariant mass is near the $Z'$ mass.
Though it depends on the size of couplings, typically we need only $L \sim 1 ~\fb^{-1}$ (to see 10 events of the dilepton resonance) for $M_{Z'} \sim 1 ~\tev$ \cite{Lee:2010hf}, which satisfies the Tevatron limit on $M_{Z'}$ \cite{Aaltonen:2008vx,Aaltonen:2008ah} for a given coupling.

For some other models where the flavor-dependent $Z'$ resonances were discussed, see Refs.~\cite{Chen:2008za,Chen:2009fx,Salvioni:2009jp,Lee:2010hf}.

\begin{table*}[tb]
\begin{tabular}{|c|l|l|l|}
\hline
~~$x_1$, $x_2$, $x_3$~~ & ~LSP~ & ~2-lepton $\til\nu$ LSP resonance~ & ~4-lepton $Z'$ resonance (via $\til\nu$ LSP)~ \\
\hline\hline
\multirow{3}{*}{$0$, $0$, $3$}
& ~$\til\nu_e$ ($z[\nu_e] = 0$)         & ~$\til\nu_e \to e\mu, \mu\mu$  & ~forbidden \\
& ~$\til\nu_\mu$ ($z[\nu_\mu] = 0$)     & ~$\til\nu_\mu \to ee, e\mu$    & ~forbidden \\
& ~$\til\nu_\tau$ ($z[\nu_\tau] \ne 0$) & ~forbidden                     & ~$Z' \to \til\nu_\tau \til\nu^*_\tau \to (\tau ~\text{included})$~ \\
\hline
\multirow{3}{*}{$0$, $3$, $0$}
& ~$\til\nu_e$ ($z[\nu_e] = 0$)         & ~$\til\nu_e \to \mu\mu$        & ~forbidden \\
& ~$\til\nu_\mu$ ($z[\nu_\mu] \ne 0$)   & ~forbidden                     & ~$Z' \to \til\nu_\mu \til\nu^*_\mu \to e\mu + e\mu$~ \\
& ~$\til\nu_\tau$ ($z[\nu_\tau] = 0$)   & ~$\til\nu_\tau \to ee, \mu\mu$ & ~forbidden \\
\hline
\multirow{3}{*}{$3$, $0$, $0$}
& ~$\til\nu_e$ ($z[\nu_e] \ne 0$)       & ~forbidden                     & ~$Z' \to \til\nu_e \til\nu^*_e \to e\mu + e\mu$~  \\
& ~$\til\nu_\mu$ ($z[\nu_\mu] = 0$)     & ~$\til\nu_\mu \to ee$          & ~forbidden \\
& ~$\til\nu_\tau$ ($z[\nu_\tau] = 0$)   & ~$\til\nu_\tau \to ee, \mu\mu$ & ~forbidden \\
\hline
\multirow{3}{*}{$-3$, $6$, $0$}
& ~$\til\nu_e$ ($z[\nu_e] \ne 0$)       & ~forbidden                     & ~$Z' \to \til\nu_e \til\nu^*_e \to (\tau ~\text{included})$~ \\
& ~$\til\nu_\mu$ ($z[\nu_\mu] \ne 0$)   & ~forbidden                     & ~$Z' \to \til\nu_\mu \til\nu^*_\mu \to (\tau ~\text{included})$~ \\
& ~$\til\nu_\tau$ ($z[\nu_\tau] = 0$)   & ~$\til\nu_\tau \to ee, \mu\mu$ & ~forbidden \\
\hline
\multirow{3}{*}{$-3$, $0$, $6$}
& ~$\til\nu_e$ ($z[\nu_e] \ne 0$)       & ~forbidden                     & ~$Z' \to \til\nu_e \til\nu^*_e \to e\mu + e\mu$~ \\
& ~$\til\nu_\mu$ ($z[\nu_\mu] = 0$)     & ~$\til\nu_\mu \to ee$          & ~forbidden \\
& ~$\til\nu_\tau$ ($z[\nu_\tau] \ne 0$) & ~forbidden                     & ~$Z' \to \til\nu_\tau \til\nu^*_\tau \to (\tau ~\text{included})$ \\
\hline
\multirow{3}{*}{$0$, $-3$, $6$}
& ~$\til\nu_e$ ($z[\nu_e] = 0$)         & ~$\til\nu_e \to \mu\mu$        & ~forbidden \\
& ~$\til\nu_\mu$ ($z[\nu_\mu] \ne 0$)   & ~forbidden                     & ~$Z' \to \til\nu_\mu \til\nu^*_\mu \to e\mu + e\mu$ \\
& ~$\til\nu_\tau$ ($z[\nu_\tau] \ne 0$) & ~forbidden                     & ~$Z' \to \til\nu_\tau \til\nu^*_\tau \to (\tau ~\text{included})$ \\
\hline
\end{tabular}
\caption{Some examples of contrasting SUSY search channels at the LHC in the minimal $B_3$ model (with the $\til\nu$ LSP). Only the light lepton ($e$, $\mu$) final states are considered. It is indicated as ``forbidden'' when the $\til\nu$ LSP production is not allowed by the new $U(1)$ gauge symmetry. It is indicated as ``$\tau$-included'' when the final leptons always include $\tau$.}
\label{tab:complementarity}
\end{table*}

\vspace{0.5cm}
{\it (ii) Dilepton $\widetilde \nu$ LSP resonance at the LHC:}
Dilepton sneutrino ($\widetilde \nu$) LSP resonance is one of the typical channels for SUSY search in the absence of the $R$-parity.
The sneutrino can be produced at the LHC by $\lambda' LQD^c$ term and it can decay into two charged leptons by $\lambda LLE^c$ term (Figure~\ref{fig:diagrams} (a)).

For the simplicity, we will assume only the LSP case when we consider a sneutrino resonance.
Otherwise, the dilepton branching ratio of the sneutrino would be diluted by the decay to the LSP plus SM particles.

Under $U(1)_{B-x_i L}$, existence of some couplings depends on lepton flavors.
\begin{itemize}
\item $\widetilde \nu$ production: Presence of $\lambda'_{ijk} L_iQ_jD_k^c$ requires the sneutrino $\widetilde \nu_i$ to have $0$ charge ($z[L_i] + 1/3 - 1/3 = 0$).
\item $\widetilde \nu$ decay: Presence of $\lambda_{ijk} L_iL_jE_k^c$ requires the same charges for leptons $\ell_j$ and $\ell_k$ ($0 + z[L_j] + z[E_k^c] = 0$).
\end{itemize}
This means, unlike usual $R$-parity violating models, some $\widetilde \nu$ resonances may be forbidden in our model.

For instance, with $x_i = (-3, 6, 0)$ and $\widetilde \nu_\tau$ LSP, diagonal resonances $\widetilde \nu \to e e$ (by $\lambda_{311} = -\lambda_{131}$) and $\mu \mu$ (by $\lambda_{322} = -\lambda_{232}$) are allowed, yet off-diagonal resonances $\widetilde \nu \to e \mu$ (by $\lambda_{312} = -\lambda_{132}$) are forbidden.
Similarly, there are no $\tau e$, $\tau \mu$, $\tau \tau$ resonances in this example\footnote{
The off-diagonal dilepton $\widetilde \nu$ resonance is possible only for $x_i = (0,0,3)$ and its permutations.
This charge assignment is basically the same as Refs.~\cite{Ma:1997nq,Ma:1998dp,Ma:1998dr}.
In that case, the model can have $B_3$ under which the proton never decays.
}.

Interestingly, the allowed final states ($e^+ e^-$ and $\mu^+ \mu^-$ only) are the same as the dilepton $Z'$ resonance case in this example.
The spin of the intermediate particle ($\widetilde \nu$ or $Z'$) can be identified from the angular distribution of the leptons though. (For example, see Ref.~\cite{Buckley:2008pp} and references therein.)

The sneutrino LSP at the LHC has been studied in the literature.
(For instances, see Refs.~\cite{Moreau:2000bs,Bernhardt:2008jz,Osland:2010yg}).
In Ref.~\cite{Bernhardt:2008jz}, they obtained the sneutrino production cross section $\sigma(p p \to \til \nu + X) \sim 10^6 ~\fb$ with $\lambda' \sim 0.1$ for a choice of mSUGRA benchmark scenario.

Since some of the $\lambda'$ should exist for the sneutrino production, the $\til \nu$ LSP branching ratio to the dilepton decay through $\lambda$ is diluted by the diquark decay through $\lambda'$.
Since the bounds on the individual $\lambda$ (for specific $ijk$ values) are not severely constrained and they can be even weaker for heavier superparticle masses, it may be still possible to have relatively large $\lambda$ values that can provide considerable dilepton branching ratio though it would call for a detailed simulation to make a quantitative prediction.
For an example of the numerical study on the dilepton sneutrino resonance at the LHC, see Ref.~\cite{Osland:2010yg}.

\vspace{0.5cm}
{\it (iii) Complementarity between the $\widetilde \nu$ and $Z'$ resonances at the LHC:}
While the dilepton sneutrino LSP resonance is a good channel to search for the SUSY without $R$-parity, the sneutrino cannot be produced at the LHC unless its $U(1)$ charge is zero in our model.
What would be an alternative way to search for the SUSY in this case?

Although the $\widetilde \nu$ with nonzero $U(1)$ charge cannot be produced directly from quarks, its coupling to $Z'$ is guaranteed.
The $Z'$ can decay into a pair of the $\widetilde \nu$ LSP, and each $\widetilde \nu$ LSP can decay into 2 charged leptons through the $\lambda$ coupling (Figure~\ref{fig:diagrams} (b)).
If the $Z'$ is heavy enough, everything will be on-shell and there will be a 4-lepton resonance.
Thus, we can find a complementarity between the $\widetilde \nu$ resonance and the $Z'$ resonance in SUSY search at the LHC.
\begin{itemize}
\item Direct $\widetilde \nu$ production ($q \bar q \to \widetilde \nu$): $z[\nu] = 0$
\item $\widetilde \nu$ production via $Z'$ ($q \bar q \to Z' \to \widetilde \nu \widetilde \nu^*$)\footnote{Note that 4-leptons are produced by the $\lambda_{ijk} L_i L_j E^c_k$ terms only when they satisfy the condition~\eqref{eq:lambda} of $x_k = x_i + x_j$.
Therefore, the complementary SUSY search by the 4-lepton resonance holds only when this condition is satisfied.}: $z[\nu] \ne 0$
\end{itemize}
When Eq.~\eqref{eq:lambda} is combined with the anomaly-free condition~\eqref{eq:xSum}, it follows that a $\lambda_{ijk}$ term is allowed only when either [$x_i = 0$ and $k=j$] or [$x_j = 0$ and $k=i$] is satisfied.
It means that models without vanishing $U(1)$ charge for a lepton flavor, such as $x_i = (9,-3-3)$, would not show the aforementioned complementarity.
Then it can be derived that the $x_1 = 0$ case has only $\lambda_{122} = -\lambda_{212}$ and $\lambda_{133} = -\lambda_{313}$ as nonvanishing $\lambda$ terms.
(Similarly, the $x_2 = 0$ case with only $\lambda_{211} = -\lambda_{121}$ and $\lambda_{233} = -\lambda_{323}$ terms, and the $x_3 = 0$ case with only $\lambda_{311} = -\lambda_{131}$ and $\lambda_{322} = -\lambda_{232}$ terms.)

Table~\ref{tab:complementarity} shows some examples containing vanishing $x_i$ values in the SUSY search at the LHC with the $\til\nu$ LSP.
It contrasts the dilepton $\til\nu$ LSP resonance and the 4-lepton $Z'$ resonance.

The 4-lepton $Z'$ resonance was first studied in Ref.~\cite{Lee:2008cn} in the family-independent charge case for the MSSM sector with exotic quarks.
The complementarity between the $\widetilde \nu$ and $Z'$ resonances is absent there though because $\lambda$ and $\lambda'$ are switched on/off by a $U(1)$ only simultaneously in the family-independent case.
The numerical result of Ref.~\cite{Lee:2008cn} indicates, however, that the 4-lepton resonance signal at the LHC can be quite sizable.

The 6-lepton $Z'$ resonance discussed in Ref.~\cite{Barger:2009xg} is absent in the $B-x_i L$ model since Higgs doublets have zero $U(1)$ charges.

\section{Baryon tetrality for the four fermion generations case}
\label{sec:B4}
In this section, we turn our attention to the 4th-generation fermion scenarios.
Recently, there has been ample interest in the 4th-generation models in many aspects including the electroweak precision test \cite{Kribs:2007nz,Chanowitz:2009mz,Erler:2010sk}.
In this view, it would be appropriate to extend our discussion to the possible 4th-generation scenarios.

Following the procedure described in Section~\ref{sec:conditions}, it is straightforward to see what are the conditions to have the minimal $B-x_i L$ and the counterpart of $B_3$ residual discrete symmetry for the 4th-generation scenario.
The conditions similar to Eqs.~\eqref{eq:xSum}, \eqref{eq:zS}, \eqref{eq:integer} are
\bea
\!\!\!\!\!\!\! && x_1 + x_2 + x_3 + x_4 = 4 ~~ \text{(anomaly cancellation)} \label{eq:xSum4}, \\
\!\!\!\!\!\!\! && z[S_1] = -z[S_2] = 4 ~~ \text{(total $Z_4$ symmetry)}, \label{eq:zS4} \\
\!\!\!\!\!\!\! && x_i = 4 \times \text{integer} ~~ \text{($Z_4 = B_4$)}, \label{eq:integer4}
\eea
respectively, in the 4th-generation scenario.

We shall call the resulting discrete symmetry $B_4$ or {\it baryon tetrality}, which has the discrete charge of
\be
B_4 :~ q = - B + 2 Y \md 4 .
\ee
It has a selection rule of $\Delta B = 4 \times \text{integer}$, which forbids proton decay ($\Delta B = 1$ process).

While it is possible to have the $B_3$ and $R$-parity from the same gauge origin with $Z_6 = B_3 \times R_2$ \cite{Lee:2010hf}, it is not possible to have the $B_4$ and $R$-parity from the same gauge origin.
It is because $Z_N = Z_n \times Z_m$ requires $n$ and $m$ to be coprime to each other.
Since it is rather straightforward to extend the LHC phenomenology of Section~\ref{sec:LHC} to the 4th-generation case, we will not discuss it in this paper.

Though we have discussed the $B_3$ and $B_4$ in the SUSY framework, the absolute proton stability under these symmetries does not require the SUSY.
In Ref.~\cite{Lee:2011jk}, $B-4L_4$ gauge symmetry was suggested as an auxiliary symmetry of the 4th-generation scenario in the non-SUSY framework.
There, it demanded 4th-generation Majorana neutrino mass term $SN_4^cN_4^c$ in order to address the dark matter issue with the 4th-generation right-handed neutrino, which cannot satisfy Eq.~\eqref{eq:zS4}.

If we do not consider the 4th-generation right-handed neutrino as a dark matter candidate, however, $U(1)_{B-4L_4}$ can satisfy all conditions~\eqref{eq:xSum4}, \eqref{eq:zS4}, \eqref{eq:integer4} and have the $B_4$ as a residual discrete symmetry.
Further, since the $B-4L_4$ does not particularly require the existence of the right-handed neutrinos for the three SM generations while having a heavy neutrino, the mechanism of Ref.~\cite{Babu:1988ig} might be able to produce the necessary neutrino masses without introducing the three SM right-handed neutrinos.

\section{Conclusion}
\label{sec:conclusion}
We presented the minimal supersymmetric model with an extra $U(1)$ gauge symmetry containing the baryon triality or $B_3$ as its only remnant discrete symmetry.
With $B_3$, proton stability is guaranteed without $R$-parity, but the lepton number violating interactions are allowed at the tree level.

Demand of the minimal particle content requires $U(1)$ charges to be family-dependent.
This leads to a flavor-dependent phenomenology at the LHC.
Interestingly, there is a complementarity between the 2-lepton resonance and the 4-lepton resonance in the SUSY search when a condition for the relevant $\lambda$ to exist is satisfied.
Thus, this model suggests a connection between the proton stability and the flavor physics in the SUSY framework.

We also extended our discussion to the four fermion generations case and introduced the baryon tetrality or $B_4$, which would play a similar role as the $B_3$ in the 4th-generation case.

\section*{Acknowledgments}
I am grateful to C. Luhn and E. Ma for helpful discussions over a long period of time.
This work was supported by US Department of Energy Grant No.~DE-AC02-98CH10886.



\end{document}